\documentclass{article}

\def\fnote#1#2{\begingroup\def\thefootnote{#1}\footnote{#2}
    \addtocounter{footnote}{-1}\endgroup}
\def\sato{mmsato11@yahoo.co.jp}

\usepackage[dvipdfmx]{graphicx}

\usepackage{amsmath}
\usepackage{amsfonts}

\begin{document}

\pagestyle{empty}
\vspace{16pt}
\begin{center}
{\Large \bf Kernel/Metric Correspondence of Dissipative Systems in Information Theory}\\
\vspace{16pt}


Masamichi Sato\fnote{*}{\sato}
\vspace{16pt}

{\sl Dai Nippon Printing, Co., Ltd.}
\vspace{12pt}\\
{\bf ABSTRACT}
\vspace{12pt}\\

\begin{minipage}{4.8in}
We study the kernel/metric correspondence pointed out in our previous work in a dissipative system which is accompanying fractional Brownian motion.
We also give some comments on information causality.
\end{minipage}

\end{center}
\vfill
\pagebreak

\pagestyle{plain}
\setcounter{page}{1}

\baselineskip=16pt

\section{Introduction}

Referring to Amari's seminal works of information geometry~\cite{amari0, amari}, we have seen an analogue of the AdS/CFT correspondence in information theory~\cite{sato0}.

Following our previous work~\cite{sato0}, we study the aforementioned correspondence for the system which consists of stochastically deviating data.
The stochastic fluctuation treated in this study follows fractional Brownian motion.
We study the kernel/metric correspondence when accompanying with this fluctuation.
We also some comments on causality.

This paper is organized as follows:
Section two is a short review of kernel in information geometry.
Section three is a description of dissipative system.
In section four, we see the relationship between stochastic fluctuation and kernel.
In section five, we consider the case of fractional Brownian motion.
In section six, we give some comments on causality.
Section seven is conclusions.

\section{Kernel in information geometry}

At first, we give the relationship between metric of data space and kernel.
This is represented as~\cite{amari}

\begin{equation}
g_{ij}=\left . \frac{\partial}{\partial x_i}\frac{\partial}{\partial x'_j}K(x_i,x'_j)\right |_{x_i=x'_j}
\end{equation}

\noindent
Here, $g_{ij}$ is the metric of data space, $x$ is point and $K$ is the kernel.

\section{Description of dissipative system}

Here, we give a short review of Fick's law~\cite{natsuume}.
For static state, the flux is proportional to gradient of density.
This is described as

\begin{equation}
J=-D\frac{\partial c}{\partial x}
\end{equation}

\noindent
Here, $J$ is flux, $D$ is diffusion coefficient, and $c$ is density.
This is called as Fick's first law.
For non-static state, the following equation holds:

\begin{equation}
\frac{\partial c}{\partial t}=-{\rm div} J=D\nabla ^2 c
\end{equation}

\noindent
This is diffusion equation and is equivalent to the equation of continuity.
This is called as Fick's second law.

\section{Stochastic fluctuation and kernel}

Consider the case of stochastic fluctuation of point, which follows standard Brownian motion:

\begin{equation}
dx=\sigma xdz.
\end{equation}

\noindent
Bellman equation for kernel derives the diffusion equation:

\begin{equation}
\frac{\partial K}{\partial t}-\frac {1}{2}\sigma ^2 \frac{\partial ^2}{\partial x^2}K=0
\end{equation}

\noindent
The solution of this equation (with the boundary condition $K(\infty) =0)$ is,

\begin{equation}
K=\frac{1}{2\sqrt{\pi \sigma^2 t}}\exp(-\frac{x^2}{4\sigma ^2 t}).
\end{equation}

\noindent
This is the derivation of Gaussian kernel~\cite{sato}.

\section{The case of fractional Brownian motion}

\subsection{Fractional Brownian motion and kernel on boundary}

Next, we consider the case of fractional Brownian motion,

\begin{equation}
d\hat{x}=\sigma \hat{x}dz^H.
\end{equation}

\noindent
Here, $H$ is called as Harst constant and we assume $H\in (\frac{1}{2},1)$.
Infinitismal differentiation of kernel derives,

\begin{equation}
d\hat{K}=\sigma \hat{x} \frac{\partial \hat{K}}{\partial \hat{x}}(dz^H-Ht^{2\alpha}dt)+\sigma^2Ht^{2\alpha}\hat{x}^2\frac{\partial^2 \hat{K}}{\partial \hat{x}^2})dt, 
\end{equation}

\noindent
here, $\alpha=H-\frac{1}{2}$.
From this equation, we obtain,

\begin{equation}
\frac{\partial \hat{K}}{\partial t}-\sigma^2Ht^{2\alpha}\hat{x}^2\frac{\partial^2 \hat{K}}{\partial \hat{x}^2}=0
\end{equation}

\noindent
Substituting $\tau=t^{2H}$ and setting

\begin{equation}
\hat{y}\equiv \frac{1}{\sigma}\left ( \log \hat{x} -\frac{\sigma^2}{2}t^{2H}\right ),
\end{equation}

\noindent
derive

\begin{equation}
\frac{\partial u(\tau, \hat{y})}{\partial \tau}-\frac{1}{2}\frac{\partial^2 u(\tau, \hat{y})}{\partial \hat{y}^2}=0.
\end{equation}

\noindent
This is the diffusion equation.
The solution of this equation is given as, 

\begin{equation}
u(\tau, \hat{y})=\frac{1}{\sqrt{2\pi \tau}}\exp(-\frac{\hat{y}^2}{2\tau}).
\end{equation}

\noindent
This solution can be interpreted as the kernel for data fluctuating with fractional Brownian motion.

\subsection{Metric of bulk in dissipative kernel/metric correspondence}

We consider the stochastic deviation from a definite space which follows fractional Brownian motion~\cite{fractional1},

\begin{equation}
d\tilde{x}_i=dx_i+d\hat{x}_i.
\end{equation}

\noindent
The metric with this deviation becomes,

\begin{eqnarray}
d\tilde{s}^2&=&ds^2+g_{ij}\{ \sigma_i \sigma_j dz_i{H}dz_j^{H}+\sigma_i\sigma_j(dx_idz_j^{H}+dx_jdz_i^{H})\}\nonumber \\
&=&ds^2+g_{ij}\sigma_i \sigma_j dz_i^{H}dz_j^{H}+o(dt^{1+H})\nonumber \\
&\sim&ds^2+g_{ij}\sigma_i \sigma_j dt^{2H}{\bf e}^i{\bf e}^j.\nonumber
\end{eqnarray}

\noindent
Here, ${\bf e}_i=\frac{dx_i}{|dx_i|}$.
We take $H=1-\epsilon$ and expand for small $\epsilon$,

\begin{eqnarray}
dt^{2H}&=&dt^{2(1-\epsilon)}\nonumber \\
&\sim&dt^2(1-2 \epsilon \ln (dt)).
\end{eqnarray}

\noindent
This gives,

\begin{equation}
d\tilde{s}^2\sim g_{\mu\nu}dx^{\mu}dx^{\nu}+g_{ij}\sigma_i \sigma_j\{1- 2 \ln (dt)\epsilon\}dt^2{\bf e}^i{\bf e}^j.
\label{met}
\end{equation}

\noindent
Next, we derive the metric from kernel

\begin{eqnarray}
\hat{g}_{ij}(\hat{x})&=&\left . \frac{\partial^2}{\partial \hat{x}_i \partial \hat{x}'_j}u(\tau, \hat{x}-\hat{x}')\right |_{\hat{x}=\hat{x}'}\\
&=&\left . \frac{\partial \hat{y}_i}{\partial \hat{x}_i}\frac{\partial \hat{y}'_j}{\partial \hat{x}'_j}\frac{\partial^2}{\partial \hat{y}_i \partial \hat{y}'_j}u(\tau, \hat{y}-\hat{y}')\right |_{\hat{y}=\hat{y}'}
\end{eqnarray}

\noindent
Here,

\begin{eqnarray}
\left . \frac{\partial^2}{\partial \hat{y}_i \partial \hat{y}'_j}u(\tau, \hat{y}-\hat{y}')\right |_{\hat{y}=\hat{y}'}&=&\frac{1}{\sqrt{\pi} (2\tau)^{3/2}}\\
&=&\frac{1}{\sqrt{2^3\pi}t^{3H}}
\end{eqnarray}

\noindent
and

\begin{equation}
\frac{\partial \hat{y}_i}{\partial \hat{x}_i}=\frac{1}{\sigma_i \hat{x}_i}.
\end{equation}

\noindent
these give the metric,

\begin{equation}
\hat{g}_{ij}(\hat{x})\sigma_i \sigma_j=\frac{1} {\hat{x}_i\hat{x}'_j \sqrt{2^3\pi}t^{3H}}.
\end{equation}

\noindent
Notice that the the order of coordinates of this equation is the same as the metric of AdS space.

If $\hat{x}_i\sim e^{-\frac{\sigma_i^2}{2}t^{2H}}$, the metric will be

\begin{eqnarray}
\hat{g}_{ij}(\hat{x})\sigma_i \sigma_j&\sim& \frac{e^{\frac{\sigma_i^2+\sigma_j^2}{2}t^{2H}}}{\sqrt{2^3\pi}t^{3H}}\\
&\sim& \frac {e^{a t^2}}{\sqrt{2^3\pi} t^3}\{1-2\left( a t^2-\frac{3}{2}\right) \ln (t) \epsilon \},
\label{met_from_kernel}
\end{eqnarray}

\noindent
here, $a=\frac{\sigma_i^2+\sigma_j^2}{2}$.
The comparison of eq.~(\ref{met}) and (\ref{met_from_kernel}) gives

\begin{equation}
t^2=\frac{5}{\sigma_i^2+\sigma_j^2}.
\end{equation}

\noindent
This gives the scale of time to which kernel/metric correspondence holds indifferent to Harst constant.

\section{Comments on causality}

The time-correlation between past and future is calculated as follows~\cite{fractional2}:

\begin{eqnarray}
C_{\Delta z^H}(t)&=&\frac{{\rm E}[(z^H(t)-z^H(t-\Delta t))(z^H(t+\Delta t)-z^H(t))]}{{\rm E}[(z^H(t)-z^H(t-\Delta t))^2]}\nonumber \\
&=&\frac{\sigma^2(2^{2H-1}-1)|\Delta t|^{2H}}{\sigma^2|\Delta t|^{2H}}\nonumber \\
&=&2^{2H-1}-1\nonumber
\end{eqnarray}

\noindent
This derived correlation becomes positive for $1/2<H<1$, zero for $H=1/2$, and negative for $0\leq H<1/2$.

The analysis of dimension to $d\hat{x}=\sigma \hat{x}dz^H$ shows that physical causality is broken for $H<1$, because the information travels in proportion to the time of order of lower than $1$.
If the broken causality is very small, the analysis with small $\epsilon$ of the last section will be effective.

\section{Conclusions}
In this paper, we studied the kernel/metric correspondence in information theory, when accompanying fractional Brownian motion.
This corresponds to a special case of our previous work.
However, this is merely one of reconfirmations of Amari's great works.

\begin{center}
    {\Large {\bf Acknowledgements} }\\
\end{center}
We greatly thank to kind hospitality of my colleagues for giving ideas and comfortable environments.
\vspace{.1in}

\pagebreak

\end{document}